\documentclass[useAMS,usenatbib]{mn2e} 
\usepackage{amssymb}
\usepackage{amsmath}
\usepackage{amsxtra}
\usepackage{graphicx}
\usepackage{rotating}
\def\aa{A\&A}

   \title[Is there a caustic crosing in Q2237$+$0305?]
   {Is there a caustic crossing in the lensed quasar Q2237$+$0305
   observational data record?}

   \author[Gil-Merino et al.]{R. Gil-Merino$^{1}$, J. Gonz\'alez-Cadelo$^{2}$,
           L.J. Goicoechea$^{2}$, V.N. Shalyapin$^{3}$ \and G.F. Lewis$^{1}$\\
             $^{1}$Institute of Astronomy, School of Physics,
              The University of Sydney, NSW 2006, Australia\\
              E-mail: [rodrigo,gfl]@physics.usyd.edu.au	 \\    
              $^{2}$Departamento de F\'{i}sica Moderna, Universidad de Cantabria
	      Avda. de Los Castros s/n, 39005 Santander, Spain\\
              E-mail: juan.gonzalec@alumnos.unican.es, goicol@unican.es \\
              $^{3}$Institute for Radiophysics and Electronics, National Academy 
              of Sciences of Ukraine, 12 Proskura St., Kharkov 61085, Ukraine\\
	      E-mail: vshal@ire.kharkov.ua \\
           }

\begin{document}

   \date{Received; accepted}
   \pagerange{\pageref{firstpage}--\pageref{lastpage}} \pubyear{2006}
   \maketitle
   \label{firstpage}

\begin{abstract}
   We re--investigate the gravitationally lensed system Q2237$+$0305 data record
   to quantify the probability of having a caustic crossing in the A component.
   Several works assume that this is the case, but no quantitative analysis is
   available in the literature.
   We combine the datasets from the OGLE and GLITP collaborations to accurately 
   trace the prominent event in the lightcurve for the A component of the system.
   Then the observed event is compared with synthetic light curves derived from 
   trajectories in magnification maps. These maps are generated using a ray--tracing 
   technique. We take more than 10$^9$ trajectories and test a wide range of different
   physical properties of the lensing galaxy and the source quasar (lens transverse 
   velocity, microlens mass, source intensity profile and source size).
   We found that around 75\% of our good trajectories (i.e. that are consistent with the
   observations) are caustic crossings. In addition, a high transverse velocity 
   exceeding 300 km s$^{-1}$, a microlens mass of about 0.1~M$_\odot$ and a small 
   standard accretion disk is the best parameter combination.
   The results justify the interpretation of the OGLE--GLITP event in Q2237+0305A as
   a caustic crossing. Moreover, the physical properties of the lens and source are in 
   very good agreement with previous works. We also remark that a standard accretion 
   disk is prefered to those simpler approaches, and that the former should be used in
   subsequent simulations.
\end{abstract}

   
%\titlerunning{Is there a caustic crosing in Q2237$+$0305?}
%\authorrunning{Gil-Merino et al.}
%
%________________________________________________________________

\section{Introduction}

The system QSO 2237$+$0305 is formed by a distant source quasar at $z_Q=1.695$ 
and a spiral lens galaxy at $z_G=0.039$ that quadruples the images of the source 
(Huchra et al. 1985). Nowadays, it is one of the best studied gravitationally lensed  
quasar systems which has been monitored by a number of groups in the last years (Corrigan 
et al. 1991, {\O}stensen et al. 1996, Vakulkik et al. 1997, Wo\'zniak et al. 2000, 
Alcalde et al. 2002, Schmidt et al. 2002). Among these monitoring campaigns, two
of them have produced excelent light curves: OGLE (Wo\'zniak et al. 2000) and
GLITP (Alcalde et al. 2002). The former collaboration covers a time frame of
several years, while the latter a few months but with a daily sampling, bad
weather or technical problems aside. Interestingly, the two datasets overlap each
other during a period of time. 

The recent datasets have been used to investigate several physical properties of 
the system. Limits on the size of the quasar emission region and the mass of the 
central black hole have been placed by Wyithe et al. (2000a), Yonehara (2001), 
Shalyapin et al. (2002), Goicoechea et al. (2003) and Kochanek (2004); the mass 
range of the microlenses in the lensing galaxy has been established by Wyithe et 
al. (2000b) and Kochanek (2004); and limits on the transverse velocity of the 
lensing galaxy have been reported by Wyithe et al. (1999) and Gil--Merino et al. 
(2005). To explain the prominent OGLE--GLITP event in the brightest component (A),
some of these authors have assumed a caustic crossing (e.g., Goicoechea et al. 
2003, Moreau et al. 2005). This interpretation of the OGLE--GLITP event in the A component (caustic
crossing) is supported by the best trajectories in terms of $\chi^2$ that appear 
in Kochanek (2004). Although Kochanek (2004) also showed other relatively good 
trajectories passing through complex magnification regions, the caustic crossing 
scenario became popular for studying the nature of the
source quasar, e.g., reconstruction of the intrinsic brightness profile or 
estimation of the size ratios (Bogdanov \& Cherepashchuk 2004, Goicoechea et al.
2004, Koptelova \& Shimanovskaya 2005).

In this contribution we merge the available data of the prominent OGLE--GLITP
event (it was monitorized by both OGLE and GLITP collaborations), building a single 
dataset for the A component. We try to find the probability of having a caustic 
crossing in this component considering different source profiles and sizes, 
microlenses masses and lens transverse velocities. In this way we both justify 
those works that assumed approximations to caustic crossings and future assumptions.

The paper is organised as follows: Sec.~\ref{merging} describes the way in which
the datasets from the different OGLE and GLITP collaborations are merged to produce 
a single set; Sec.~\ref{simulations} explains the method used to produce the 
synthetic light curves; Sec~\ref{results} establishes the criteria to compare the
synthetic light curves and the observational record, and gives the results of these
comparisons; finally, in Sec.~\ref{conclusions} we include our conclusions.

%___________________________________________________________

\section{Data Merging}\label{merging}

\begin{figure}\label{ogle+glitp}
   \centering
   \includegraphics[angle=-90,width=8.5cm]{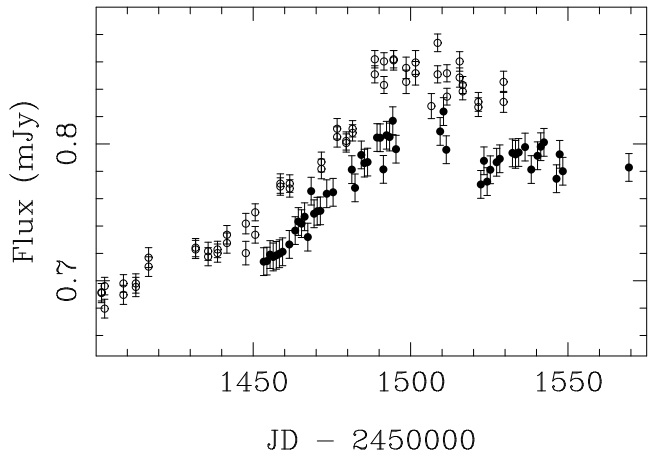}
      \caption{$V$-band photometry of the A component of Q2237$+$0305 for both data
      sets, OGLE (open circles) and GLITP (filled circles). There is still no calibration
      between the two sets.}
   \end{figure}

We are interested in the A component of Q2237$+$0305. The system was monitored
by the phase II of the OGLE\footnote{http://astrouw.edu.pl/$\sim$ogle/} project.
During part of that period, the  GLITP\footnote{data from
http://www.iac.es/proyect/gravs\_lens/GLITP/} project obtained optical frames of the
same lens system. In particular, the overlapping period was approximately between
the Julian days 2451400--2451600, corresponding to from October 1999 until February 2000. 
The corresponding observational data set from both campaings in the $V$ filter is 
displayed in Fig.~\ref{ogle+glitp}. From now on we use a simpler scale of time: 
JD--2450000, so a peak of flux around day 1500 is seen in Fig. 1.
%--------------------------------------FIGURE 1---------- 
%----------------------------------------------------------- 

In order to merge the OGLE and GLITP data sets, the first problem we found was the 
night sampling. During the OGLE campaign, two data points were consecutively obtained 
every night. The initial step is thus to ``re--reduce" these data points to get one per 
observational day. In this way, when two points are available for a given night, the 
observational date and flux will be the mean of the individual observational dates and 
fluxes. The estimation of the errors for the new points has two cases: if the individual 
error bars overlap, the error is the mean of the individual errors; if they do not 
overlap, then the error is estimated by the square root of the sum of the square of the 
average of the individual errors plus the square of the difference between the mean flux 
and one of the two individual fluxes. 

Once we have the same night sampling for both data sets, the next step is to merge them. 
To do that, we try to find a linear expression that relates the data sets:
\begin{equation}
F_{OGLE}(t)=a+b \cdot F_{GLITP}(t)
\label{eq1}
\end{equation}
where $F(t)$ are the fluxes and $a$ and $b$ are the constants to be determined. We treat 
the problem as a $\chi^2$ minimization (e.g., Ofek \& Maoz 2003, Bogdanov \& Cherepashchuk 
2004). The best fit found is 
\begin{equation}
F_{OGLE}(t)=-0.097+1.183 \cdot F_{GLITP}(t)
\label{eq2}
\end{equation}

In the GLITP photometric system, the ``new" OGLE data points are then
\begin{equation}
F'_{OGLE}(t)=0.082+0.845 \cdot F_{OGLE}(t) \nonumber
\end{equation}
\begin{equation}
\sigma'_{OGLE}=0.845 \cdot \sigma_{OGLE}
\label{eq3}
\end{equation}

\begin{figure}\label{ogle+glitp3}
   \centering
   \includegraphics[angle=-90,width=8.5cm]{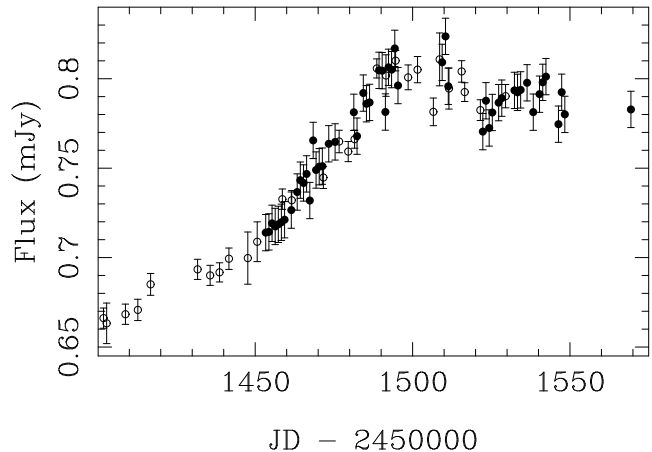}
      \caption{OGLE (open circles) and GLITP (filled circles) data sets have been merged
      following a ``re--reduction" of the original OGLE data and expressions in 
      Equation~\ref{eq3} (see text for further details).}
   \end{figure}

The resulting lightcurve is presented in Fig.~\ref{ogle+glitp3}. This is the
dataset we will compare to simulations.
%---------------------------------------FIGURE 2------------ 
%----------------------------------------------------------- 

%___________________________________________________________

\section{Simulations}\label{simulations}

Prior to produce synthetic light curves, we first build magnification maps where these
will be drawn. The 2--dimensional magnification maps are done using a ray tracing
method: light rays are back-traced from the observer to the source plane  through a
distribution of mass sitting on the lens plane (Kayser et al. 1986,  Wambsganss 1990).
The distribution of mass responds to a macromodel of the system,  that gives the two
relevant parameters: the convergence $\kappa$ or the distribution  of mass in compact
objects at the position of the beam and the shear $\gamma$ or the  influence of matter
outside the beam. In general, lightcurves of lensed quasars might contain both
microlensing and instrinsic variability. So in order to analyse microlensing alone one
should correct for the time delay and then subtract to components. However, in the
case of Q2237 the time delay is expected to be of a few hours (Vakulik et al. 2006).
In addition, in the case of the period analysed here for Q2237, the flatness of
component D indicates that intrinsic variability does not contribute significantly to
the behavior of the components (see Shalyapin et al. 2002). We use the parameters for
the A component  given by Schmidt et al (1998): $\kappa=0.36$ and $\gamma=0.40$ (we
note that in Gil-Merino et al. 2005 slightly different models where indistinguishable,
driving to the same conclusions). The resulting  density of rays at a point in the
source plane is proportional to the microlensing magnification of the source at that
position. The space scale factor is the Einstein radius, defined in the source plane
as
\begin{equation}
r_E=\left(\frac{4GM_{\rm \mu lens}}{c^2}\frac{D_s D_{ds}}{D_d}\right)^{1/2} ,
\label{re}
\end{equation}
where $M_{\rm \mu lens}$ is the mass of the microlenses and $D_i$ are angular
distances: $D_s$ is observer--source, $D_d$ observer--lens and $D_{ds}$ lens--source. 
Setting the space scale ($r_E$), the value of the microlens mass is directly related 
to the choice of the cosmological model. The magnification maps are then convolved 
with a particular intensity profile for the source, which is characterized by a 
typical radius. We must also assume a certain effective transverse velocity for the 
source (transverse displacement of the source in the source plane divided by elapsed 
time in the observer's clock) $V_t = - (1 + z_G)^{-1}(D_s/D_d)v_{t, lens}$ (see e.g. 
Gil--Merino et al. 2005 and references therein for the relationship between the 
effective transverse motions of the source and the lens). The synthetic microlensing 
light curves are obtained extracting linear trajectories across the magnification 
patterns, since once all physical parameters are fixed and knowing the observational 
baseline of 200 days, we can calculate the length of these trajectrories. We note here
that we do not take into account the additional effect of the proper motion of the
microlenses. This effect would play a significant role  in large time-scale lightcurves
but in short time-scale campaigns the magnification maps can be considered static to
this respect. Including the proper motion of microlenses, we would increase the number
of caustic crossings in any case, so the results here are conservative.

To carry out the simulations we take $H$ = 66 km s$^{-1}$ Mpc$^{-1}$. For a given 
value of $r_E$ in Eq. (4), the microlens mass depends on the matter--energy density 
of the universe. We focus on both the Einstein--de Sitter model (EdSM: $\Omega_M$ = 1,
$\Omega_{\Lambda}$ = 0) and the current concordance model (CM: $\Omega_M$ = 0.3,
$\Omega_{\Lambda}$ = 0.7), so each cosmology leads to a different microlens mass, but
the final statistics on the caustic crossings remains the same. 
Taking a particular value of $V_t$, we also have two different values of the effective
transverse velocity for the lens, one corresponding to the EdSM and another associated
with the CM. In Table~\ref{values} all the physical values are presented. We use a set
of axisymmetric sources, including brightness distributions enhanced at the centre of
the source (standard accretion disk and and Gaussian profiles) as well as a uniform
brightness distribution. Their typical radii are consistent with previous studies by
Wyithe et al. (2000a), Yonehara (2001), Shalyapin et al. (2002) and Kochanek (2004).
Wyithe et al. (2000b) obtained that the most likely value for the microlens mass is in
the range 0.01--1 M$_{\odot}$, while Kochanek (2004) used a CM and reported an 
interval 0.003--0.1 M$_{\odot}$. The values of $M_{\mu lens}$ in Table 1 (EdSM and CM)
vary between a lower limit slightly above 0.01 M$_{\odot}$ and an upper limit slightly
below 1 M$_{\odot}$, and they reasonably agree with previous work. With respect to the
lens transverse velocities, Wyithe et al. (1999) claimed that $v_{t, lens} <$ 500 km 
s$^{-1}$ (EdSM). Moreover, using an intermediate microlens mass of 0.1 M$_{\odot}$, 
Gil--Merino et al. (2005) also inferred a constraint $v_{t, lens} <$ 630 km s$^{-1}$ 
(CM). Our values of $v_{t, lens}$ in Table 1 are basically consistent with these 
dynamical constraints.

%------------------------------------------TABLE 1-------------------------
\begin{table}[tb]
\centering
\begin{tabular}{cccccc}
\hline\noalign{\smallskip}
 intensity & source      & $M_{\mu lens}$ (M$_{\odot}$) & $v_{t, lens}$ (km s$^{-1}$) \\
 profile   & radius (cm) & EdSM/CM                      & EdSM/CM \\
\noalign{\smallskip}\hline\noalign{\smallskip}
 Uniform  & 6$\cdot$10$^{14}$ & 0.05/0.04 & 100/75  \\ 
 Gaussian & 2$\cdot$10$^{15}$ & 0.10/0.08 & 300/220 \\
 Standard & 6$\cdot$10$^{15}$ & 0.6/0.5   & 600/445 \\ 
\noalign{\smallskip}\hline
\end{tabular}
\caption{We tested three different intensity profiles, source radii, Einstein radii and 
effective transverse velocities for the source. The resulting simulations were obtained
for all the possible combinations between these parameters. Here, standard profile means 
intensity profile of a Newtonian geometrically--thin and optically--thick accretion 
disk. The source radius is the typical radius of the intensity profile, so it is the
radius of the quasar for a uniform disk, but is smaller than the quasar radial size
(containing 95\% of the total brightness) in the other cases (see Shalyapin et al. 2002).
On the other hand, the values of $M_{\mu lens}$ and $v_{t, lens}$ depend on the 
cosmological model, and we present their values for the EdSM and CM with $H$ = 66 km 
s$^{-1}$ Mpc$^{-1}$ (see main text).}
\label{values}
\end{table}
%------------------------------------------------------------------------
Taking into account that each convolution gives a different map, we have
produced a total number of 108 magnification maps, 2$\times$2~$r_E$ on 2000
pixels a side each. Considering all possible combinations of the transverse
velocities, the number of time maps (by transforming the spatial axes into
time axes) is 3 times greater. The number of trajectories tested
on each magnification pattern is 10$^7$. The total CPU time for these
simulations, excluding the generation of maps, was around 720 hours on a
Pentium~4.

But could Nature be conspiring against us? It could happen that the variability
detected in the A image (see Fig.~\ref{ogle+glitp3}) is in fact a cusp passing and
what we see is a combination of a symmetric microlensing event due to the cusp plus a
gradient originated by intrinsic variability. If this is true, that gradient should
also appear in component D during the same period, because the time delays are of only
the order of hours. Since this component is flat, the only explanation is that
microlensing and intrinsic variability are mutually canceled in image D. Fortunately
we can test this; we construct a magnification map for the D component of 4500 pixels
on a side covering a physical length of 10 Einstein radii, and we then analysed the
gradient of 10$^6$ tracks of a length equal to the  200 days period of observations,
for each combination of source size and transverse velocity. We search for gradients
between a 15\% and a 23\%, the ranges associated with the gradient seen in the A
component. We found that for the biggest source size considered in this study only
0.15\% of the cases showed a gradient in that range. The percentage was slightly
bigger, 1.16\%, for the smallest source size. These results were obtained for
transverse velocities of 600 km s$^{-1}$. Considering velocities of 300 km s$^{-1}$
the percentage was 0.5\% for the smallest source size and 0.001\% for the biggest one.
Thus, we conclude that the possibility of a conspirancy between microlensing and
intrinsic variability is extremely unlikely in this work.
%___________________________________________________________

\section{Analysis and Results}\label{results}
%-----------------------------------------FIGURE 3--------- 
   \begin{figure}
   \centering
   \includegraphics[angle=0,width=6.0cm, viewport=115 45 405 335,clip]{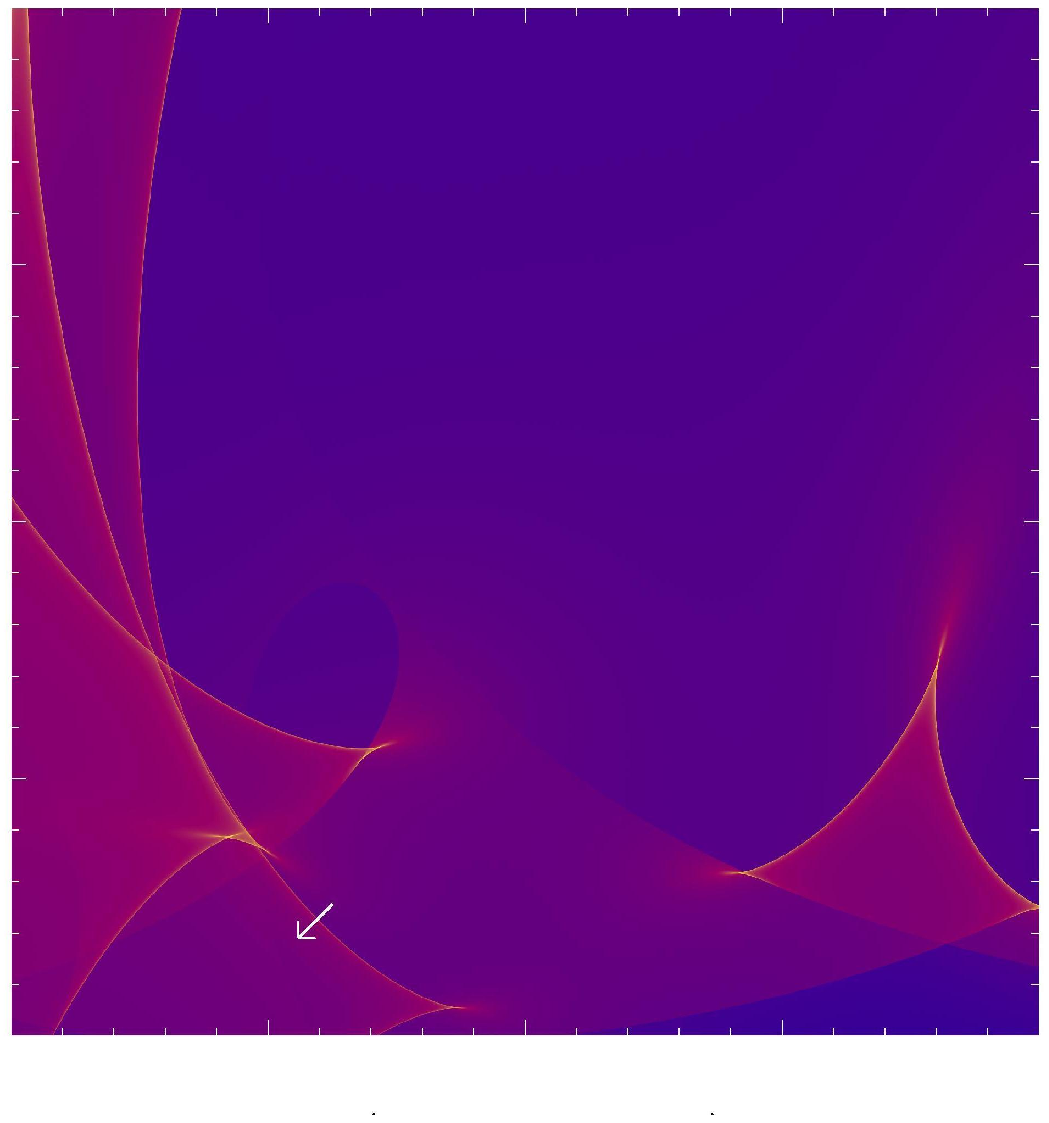}
   \includegraphics[angle=0,width=6.0cm, viewport=115 45 405 335,clip]{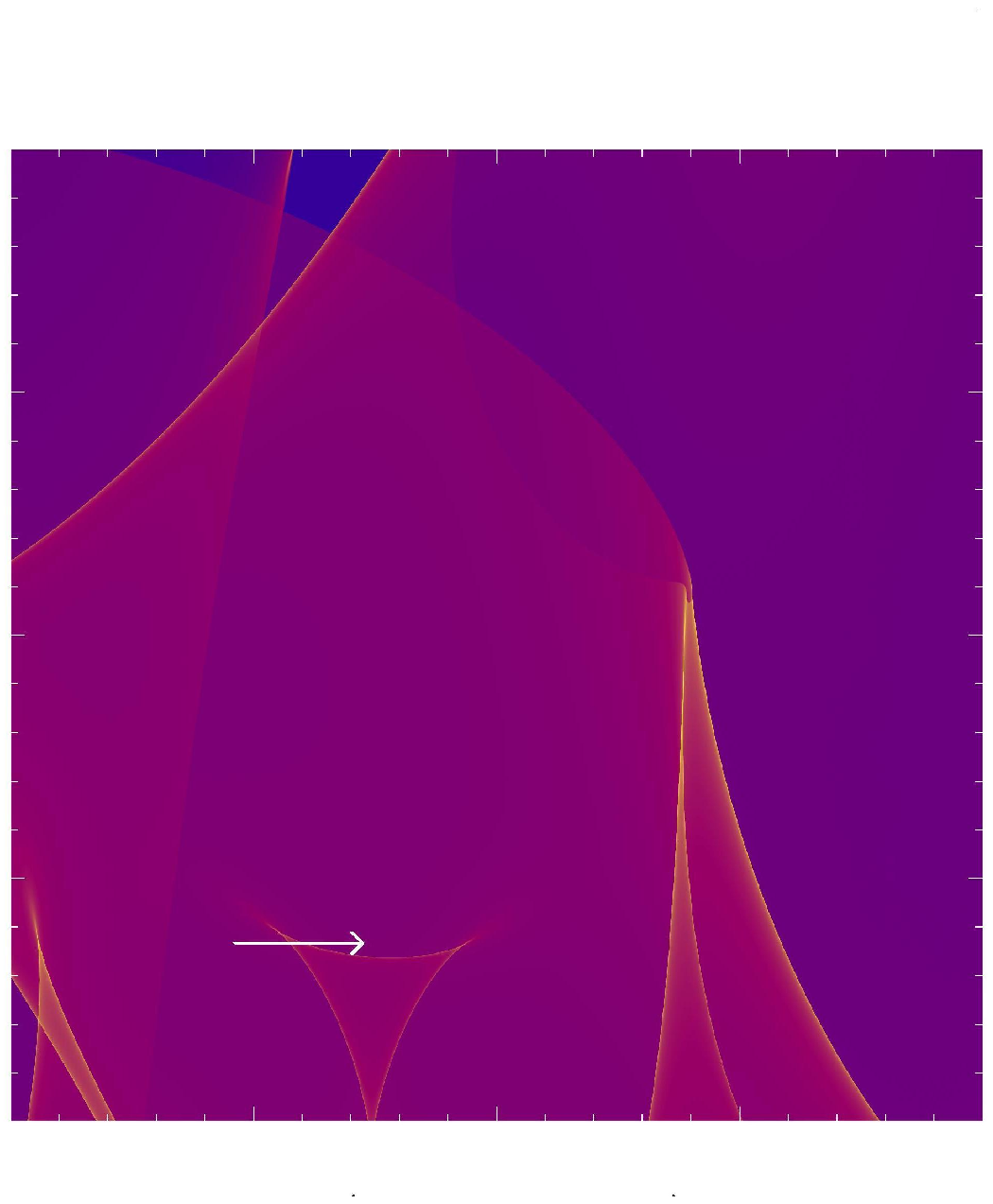}
   \includegraphics[angle=0,width=6.0cm, viewport=115 45 405 335,clip]{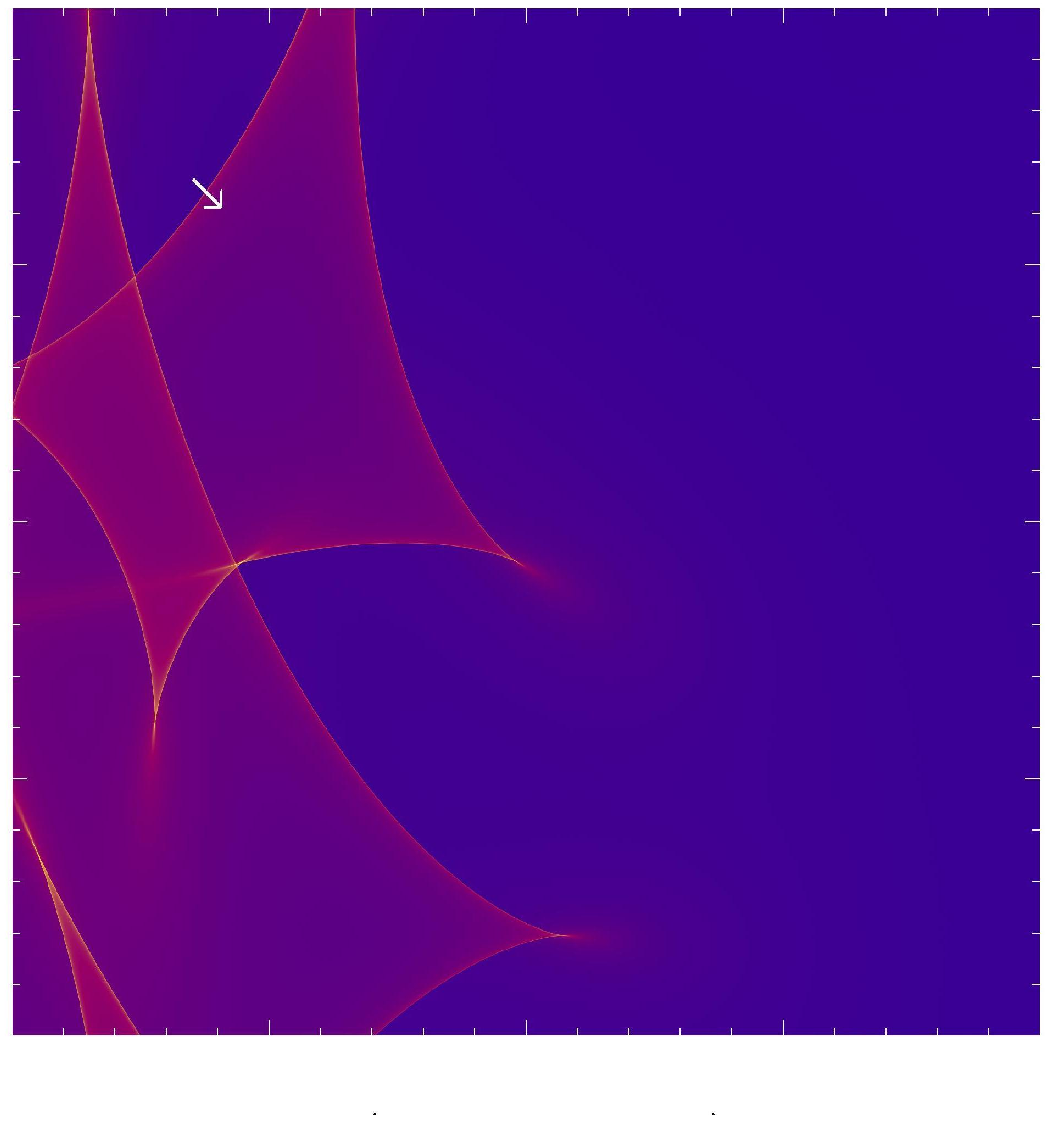}
	\caption{Caustic crossings (top and bottom panels) and cusp crossings (middle panel)
       that agree with the OGLE--GLITP event in Q2237+0305A (see Fig. 2). Their associated 
       synthetic lightcurves verify Eq.~\ref{fit}.}
         \label{scurve}
   \end{figure}
%----------------------------------------------------------- 

Each simulated lightcurve is compared to the observational one. The way we did it
is a $\chi^2$ test. We estimate the difference between each $\chi^2$ value and the
the number of degrees of freedom ($\nu$), and then we select the good fits verifying
\begin{equation}\label{fit}
\delta = \frac{\vert \chi^2-\nu \vert}{\sqrt{2\cdot \nu}} \leqslant 1
\end{equation}
where $\delta$ is the relative deviation of $\chi^2$ (Shalyapin et al. 2002). All the 
synthetic trajectories that satisfy Eq.~\ref{fit} are considered as physical processes 
consistent with the observational lightcurve, i.e., good trajectories. Then we compute 
how many of them are in fact caustic crossings in their respective magnification maps. 
After generating a total number of 3.34$\cdot$10$^9$ synthetic lightcurves, we found 
that 54 of them satisfied Eq.~\ref{fit}, with a $\langle \hat{\chi}^2 \rangle$=1.125. 
Among them, 14 were not caustic crossings, but cusp crossings. This means that the 
event seen in the A component of Q2237+0305 can be explained as a caustic crossing in 
a 74\% of the good trajectories generated in this contribution. The three panels of 
Figure 3 show some good trajectories in our magnification maps. Caustic crossings 
appear in the top and bottom panels of Fig. 3. However, the middle panel includes  
less probable (1:4) cusp crossings.

%------------------------------------------TABLE 2-------------------------
\begin{table}[tb]
\centering
\begin{tabular}{cccccc}
\hline\noalign{\smallskip}
 $M_{\mu lens}$ (M$_{\odot}$) & trajectories with $\delta \leq 1$ \\
 EdSM/CM & \\
\noalign{\smallskip}\hline\noalign{\smallskip}
0.05/0.04 & 26\% \\
0.10/0.08 & 56\% \\
0.6/0.5 & 18\% \\
\noalign{\smallskip}\hline
\end{tabular}
\caption{More than a half of the good trajectories suggests a mass of the
microlenses of about 0.1~M$_{\odot}$, among the values tested.}
\label{mass}
\end{table}
%------------------------------------------------------------------------
%------------------------------------------TABLE 3-------------------------
\begin{table}[tb]
\centering
\begin{tabular}{cccccc}
\hline\noalign{\smallskip}
 $v_{t, lens}$ (km s$^{-1}$) & trajectories with $\delta \leq 1$ \\
 EdSM/CM & \\
\noalign{\smallskip}\hline\noalign{\smallskip}
100/75 &  0\% \\
300/220 & 13\% \\
600/445 & 87\% \\
\noalign{\smallskip}\hline
\end{tabular}
\caption{The maximum transverse velocity tested is favored by the simulations against lower
values. This is interesting because the previous strongest upper limits for the effective 
transverse velocity were of $\sim$ 600 km s$^{-1}$.}
\label{vt}
\end{table}
%------------------------------------------------------------------------

Interestingly, the results can also be sorted attending to different criteria.
For example, all the trajectories matching Eq.~\ref{fit} correspond to a standard 
accretion disk. None of the other intensity profiles seem to be viable to explain
the prominent event. Due to the size of the maps, we cannot properly test the extreme
case of a standard source with the larger radius. However, considering the other two
radii, the smaller one is clearly favoured (91\% of the trajectories with $\delta 
\leq 1$). In addition, more than a half of the trajectories with $\delta \leq 1$ 
correspond to a microlens mass of about 0.1~M$_{\odot}$ (Table~\ref{mass}), and 87\% 
of them to transverse velocities of 450--600 km s$^{-1}$ (Table~\ref{vt}). 

%__________________________________________________________________

\section{Conclusions}\label{conclusions}

Several previous works using either OGLE or GLITP (or both) data sets assumed a
caustic crossing occurring in the A component of Q2237+0305 (see, e.g., Shalyapin et
al. 2002). In this contribution we tested, via ray-tracing simulations, whether the 
event seen in component A running from October 1999 to February 2000 is compatible 
with a caustic crossing and to which extend. From a very large number of simulated 
trajectories ($>$ 10$^9$) we found that only several tens were compatible with 
observations. These good trajectories are related to two different kinds of physical
scenarios: caustic crossing (see the top and bottom panels of Fig. 3) and cusp 
crossing (see the middle panel of Fig. 3). A caustic crossing scenario has a 
probability of about 75\% (40 good trajectories), whereas a cusp crossing scenario (do
not confuse with a source passing close, but outwardly, to a cusp) has a smaller
probability of about 25\% (14 good trajectories). We checked different lens 
transverse velocities with an upper limit of 600 km s$^{-1}$ (Wyithe et al. 1999, 
Gil--Merino et al. 2005). Surprisingly, small or intermediate velocities ($\leq$  300
km s$^{-1}$) do not seem to be the most probable ones, which means that the true
transverse motion might be close to that upper limit. Our results also suggested  that
the mass of the microlenses at the location of image A should be very close to 
0.1~M$_\odot$, which is in the centre of the range derived by Wyithe et al. (2000b)
and is the upper limit reported by Kochanek (2004). Yonehara (2001) and Gil--Merino 
et al. (2005) also assumed this mass to obtain their strongest constraint on the 
quasar size and the lens transverse velocity, respectively. Interesting is also the 
fact that the best results were obtained using a standard accretion disk as intensity 
profile and that the smallest source size tested was the favored one. This last 
result is also in agreement with some previous work (Shalyapin et al. 2002, Kochanek
et al. 2004). We note that we did not investigated different inclinations of the disk
nor more sophisticated accretion disk models, which is an interesting issue considered
as a second order effect but to be addressed in a future work. Very recently,
Mortonson et al. (2005) found that the generic  microlensing fluctuations are
relatively insensitive to all properties of the source  models except the half-light
radius of the disk. Here and in other related papers it is showed the feasibility of a
discrimination between different intensity profiles  through the observation and
analysis of caustic/cusp crossings.

\section*{Acknowledgements}
We thank the anonymous referee for calling our attention on the Nature conspirancy.
This research was partially supported by the Spanish Department of Education and
Science  grant AYA2004-08243-C03-02. We acknowledge support by the European
Community's Sixth  Framework Marie Curie Research Training Network Programme,
Contract  No.MRTN-CT-2004-505183 ``ANGLES".

\end{document}